\documentclass[twocolumn,showpacs,preprintnumbers,amsmath,amssymb,prd]{revtex4-1}
\usepackage{graphicx}
\usepackage{amsmath,amssymb}
\usepackage{bm}
\usepackage{hyperref}
\usepackage{xcolor}
\usepackage{booktabs}

\newcommand{\Msun}{M_\odot}

\newcommand{\kpc}{\mathrm{kpc}}
\newcommand{\xHI}{\bar{x}_{\rm HI}}
\newcommand{\sigm}{\sigma/m}
\newcommand{\dchi}{\Delta\chi^2}
\newcommand{\fstar}{f_{\star,0}}
\newcommand{\sigUV}{\sigma_{\rm UV}}
\newcommand{\alow}{\alpha_{\rm lo}}
\newcommand{\Dbind}{\Delta_{\rm bind}}
\newcommand{\Wg}{W_g}

\begin{document}

\title{Breaking the UV Luminosity Function Degeneracy: \\Self-Interacting Dark Matter Constraints from Reionization Topology}

\author{Zihan Wang}
\thanks{zihan.wang@queens.ox.ac.uk}
\affiliation{Department of Physics, University of Oxford, Oxford OX1 3PU, UK}
\author{Huanyuan Shan}
\thanks{hyshan@shao.ac.cn}
\affiliation{Shanghai Astronomical Observatory, Chinese Academy of Sciences, Shanghai 200030, China}

\date{\today}

\begin{abstract}
Self-interacting dark matter (SIDM) is the leading framework resolving small-scale cold dark matter (CDM) crises, yet high-redshift SIDM constraints are fundamentally limited by degeneracies between dark matter microphysics and galaxy formation astrophysics. We demonstrate that the UV luminosity function alone cannot constrain SIDM: star formation suppression from SIDM halo core formation is fully absorbed by modest adjustments to standard astrophysical parameters.
We show that 21 cm reionization topology breaks this degeneracy completely, providing a nuisance-immune probe: the SIDM-enhanced duty cycle of ionizing photon escape leaves a morphological signature fully independent of star formation efficiency. Combining JWST UVLF measurements with SKA1-Low forecasts, constant-cross-section SIDM with $\sigma/m \gtrsim 1$–$2\ \mathrm{cm^2/g}$ is either excluded or detectable across all physically motivated star formation coupling strengths. Our results establish a robust new avenue to probe dark matter microphysics in the early Universe.
\end{abstract}

\maketitle

\section{Introduction}

Self-interacting dark matter (SIDM), in which dark matter particles scatter with cross-section per unit mass $\sigm \sim 1$--$10$\,cm$^2$\,g$^{-1}$, remains the leading framework for resolving persistent small-scale tensions in CDM: the core--cusp problem \cite{deBlok2010,Oh2011,Moore1994}, the too-big-to-fail problem \cite{BoylanKolchin2011,Klypin1999}, and the diversity of rotation curves \cite{Kamada2017,Ren2019,Spergel2000,Tulin2018,Kaplinghat:2015aga}. All existing constraints derive from low-redshift systems: cluster mergers and lensing require $\sigm \lesssim 1$\,cm$^2$\,g$^{-1}$ at $v \sim 1000$\,km\,s$^{-1}$ \cite{Markevitch2004,Randall2008,Peter2013,Harvey2015}, while the cross-section at dwarf scales ($v \sim 30$--$50$\,km\,s$^{-1}$) remains poorly constrained.

The high-redshift Universe offers a qualitatively different testing ground. At $z > 6$, halos are less concentrated than at $z = 0$, baryonic potentials are shallower, and gas fractions are higher, amplifying the observable consequences of SIDM core formation. JWST has extended galaxy detection to $z > 14$ and measured the UV luminosity function with unprecedented precision \cite{2024MNRAS.533.3222D,2025ApJ...980..138H,kreilgaard2026beaconjwstnircampureparallel}, making the UVLF a natural target for dark matter constraints. At the same time, forthcoming SKA observations of the 21\,cm signal from the epoch of reionization will map the morphology of ionized regions on scales directly sensitive to the source population \cite{Wang:2026uuo}.

SIDM core formation modifies both observables through the same underlying physics: a reduction in the central gas binding energy $\Wg$ of high-redshift halos. Self-scattering thermalizes the inner halo within a radius $r_1$ that depends on $\sigm$, replacing the NFW cusp with a constant-density core and reducing $\Wg$ by 30--90\% for $\sigm = 1$--$10$ at $M \sim 10^{10}$--$10^{11}\,\Msun$ (Supplemental Material \S A). This single modification drives two independent observable effects. First, reduced $\Wg$ makes supernova-driven blowout more likely, increasing the duty cycle of ionizing-photon escape and enhancing the 21\,cm topology \cite{Wang:2026uuo}. Second, reduced $\Wg$ suppresses the star formation efficiency, dimming galaxies and suppressing the UVLF. The two channels are physically independent: one governs how photons escape, the other governs how many are produced.

In this Letter, we demonstrate that the UVLF channel is fundamentally limited by a complete degeneracy with astrophysical parameters, while the topology channel is immune to this degeneracy. Combining both probes yields the first robust multi-probe constraint on SIDM at high redshift.

\section{Model and Data}
\label{sec:framework}

We parametrize the SIDM modification of the SFE as
\begin{equation}
f_\star^{\rm SIDM}(M) = f_\star^{\rm CDM}(M) \times \bigl[1 - \eta\,\Dbind(M,\sigm)\bigr],
\label{eq:fstar_sidm}
\end{equation}
where $\Dbind \equiv 1 - \Wg^{\rm SIDM}/\Wg^{\rm CDM}$ is the fractional binding-energy reduction and $\eta \in [0,1]$ is the coupling efficiency between the dark matter potential and the interstellar medium. The binding-energy reduction is governed by the thermalization radius $r_1$, within which self-scatterings flatten the NFW cusp into a constant-density core. For an NFW halo,
\begin{equation}
r_1 \simeq 0.8\;\kpc \left(\frac{\sigm}{1\;\mathrm{cm^2\,g^{-1}}}\right)^{\!0.55}\!\left(\frac{M}{10^{10.5}\Msun}\right)^{\!0.35}\!\left(\frac{8}{1{+}z}\right)^{\!0.5},
\label{eq:r1}
\end{equation}
calibrated against Paper~I estimates \cite{Wang:2026uuo} (Supplemental Material \S A for the full derivation). Halos with $r_1 < r_{\rm inner} = 0.3$\,kpc have $\Dbind = 0$ and are unaffected, giving the SIDM suppression a characteristic mass dependence.

The CDM baseline adopts a double power-law SFE \cite{Mason:2015cna,Wechsler:2018pic} with four free parameters $\bm{\theta}_{\rm astro} = \{\fstar, \alow, \sigUV, z_{\rm evol}\}$: peak efficiency $\fstar$, low-mass slope $\alow$, UV magnitude scatter at fixed halo mass $\sigUV$, and redshift evolution index. The UVLF is computed by convolving the Sheth--Tormen halo mass function \cite{Sheth:1999mn} with the UV magnitude mapping using the Kennicutt relation \cite{Kennicutt:1998zb} and Gaussian scatter $\sigUV$ \cite{Mason:2015cna,2024MNRAS.533.1111E} (Supplemental Material \S B). The SIDM model adds $\bm{\theta}_{\rm DM} = \{\sigm, \eta\}$.

We fit 31 JWST data points at $z = 9$--$14$, verified against the published tables of Donnan et al.\ \cite{2024MNRAS.533.3222D} (29~points from PRIMER+JADES+NGDEEP, 2548 galaxies over $\sim 370$~arcmin$^2$) and Harikane et al.\ \cite{2025ApJ...980..138H} (2~spectroscopic points at $z \sim 14$, including GS-z14-0 at $z_{\rm spec} = 14.32$). The CDM best fit gives $\fstar = 0.019$, $\alow = 2.14$, $\sigUV = 0.65$\,mag, with $\chi^2/\mathrm{dof} = 20.3/27 = 0.75$, consistent with the JWST stellar mass function \cite{2024MNRAS.533.1808W} and abundance matching at $z = 0$ \cite{2019MNRAS.488.3143B} (Supplemental Material Fig.~S1).

For the topology, the duty-cycle enhancement is modeled as $p_{\rm SIDM} = p_{\rm CDM}(\Wg^{\rm CDM}/\Wg^{\rm SIDM})^{0.7}$ \cite{Wang:2026uuo}, and the SKA1-Low forecast is calibrated to the Paper~I result that SIDM with $\sigm = 10$ yields $\sim 15\sigma$ cumulative significance at a 5\% systematic floor in 1000\,h. The topology signal depends only on $\sigm$ through the duty cycle and is independent of $\eta$ and $\fstar$ at fixed $\xHI$, because the ionizing efficiency $\zeta$ is re-tuned to match the same neutral fraction regardless of the SFE normalization.

\section{The UVLF Degeneracy}
\label{sec:degeneracy}

We first compute the conditional $\dchi$ with all astrophysical parameters held at the CDM best fit. The log-likelihood is
\begin{equation}
\ln\mathcal{L} = -\frac{1}{2}\sum_{i=1}^{31}\left[\frac{\log_{10}\Phi_{\rm mod}(M_{{\rm UV},i}, z_i) - \log_{10}\Phi_{{\rm obs},i}}{\sigma_i}\right]^2,
\label{eq:loglik}
\end{equation}
with symmetrized errors $\sigma_i$. The SIDM-induced suppression produces large conditional $\dchi$ values: from 1.9 at $(\sigm = 0.5,\, \eta = 0.05)$ to 3093 at $(\sigm = 5,\, \eta = 0.50)$. Nominal 95\% CL limits range from $\sigm < 0.11$ at $\eta = 0.50$ to $\sigm < 0.92$ at $\eta = 0.05$ (Supplemental Material Table~S1). These conditional limits, however, assume perfect knowledge of the astrophysical parameters.

The proper statistical treatment is a profile likelihood, in which the nuisance parameters are re-optimized at each grid point:
\begin{equation}
\dchi_{\rm prof}(\sigm,\eta) = -2\Bigl[\max_{\bm{\theta}_{\rm astro}} \ln\mathcal{L}(\sigm,\eta,\bm{\theta}_{\rm astro}) - \ln\mathcal{L}_{\rm CDM}\Bigr].
\label{eq:profile}
\end{equation}
We evaluate this on an $18 \times 14 = 252$ point grid spanning $\sigm = 0$--$20$ and $\eta = 0.01$--$1.0$, using Nelder--Mead optimization at each grid point.

\begin{figure*}
\includegraphics[width=\textwidth]{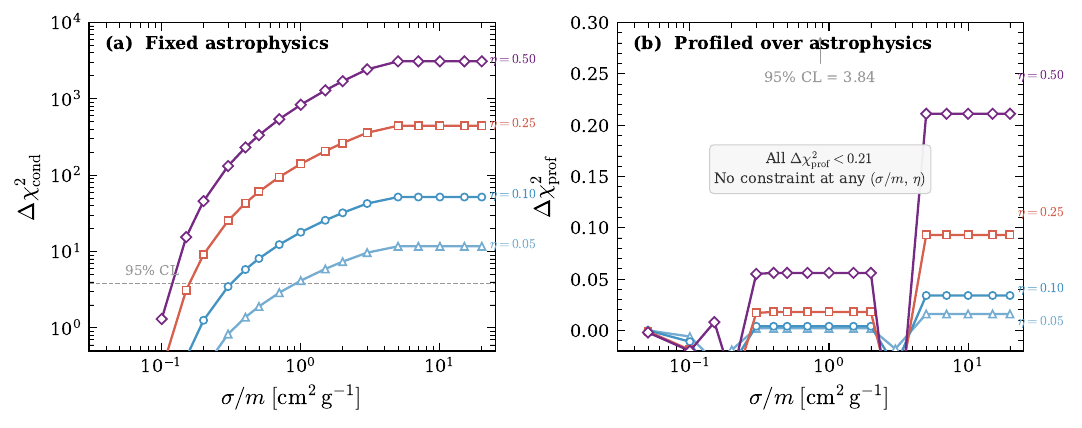}
\caption{The UVLF cannot constrain SIDM. (a)~Conditional $\dchi$ as a function of $\sigm$ at four values of the SFE coupling $\eta$, with astrophysics fixed at the CDM best fit. The dashed line marks the 95\% CL threshold ($\dchi = 3.84$). (b)~Profile likelihood $\dchi_{\rm prof}$ after re-optimizing all four astrophysical parameters at each of 252 grid points. The signal collapses to $< 0.21$ everywhere. The optimizer absorbs the SIDM suppression by increasing $\fstar$ (normalization) and broadening $\sigUV$ (shape), both within current observational uncertainties.}
\label{fig:degeneracy}
\end{figure*}

The profiled $\dchi$ collapses to $0.002$--$0.21$ across the entire grid (Figure~\ref{fig:degeneracy}), far below the 95\% CL threshold of 3.84. The optimizer compensates through two channels. It increases $\fstar$ by 12--136\%, shifting the entire UVLF brighter to absorb the normalization change from SIDM. It simultaneously broadens $\sigUV$ by 0.01--0.12\,mag, smearing out the mass-dependent shape signature that would otherwise distinguish the SIDM prediction from CDM. At $\sigm = 2$ and $\eta = 0.25$, for instance, $\fstar$ increases from 0.019 to 0.034 ($+77\%$) and $\sigUV$ from 0.65 to 0.72\,mag, recovering $\dchi_{\rm prof} = 0.018$ from a conditional value of 264 (Supplemental Material Fig.~S3). Both shifts are well within observational uncertainties: the survey-to-survey variation in the $z \sim 10$ UVLF normalization alone is a factor of $\sim 2$ between photometric and spectroscopic determinations.

Although the profiled fit formally absorbs the SIDM signal, the required $\fstar$ shifts can be tested against the stellar mass function (SMF). Weibel et al.\ \cite{2024MNRAS.533.1808W}, using JWST PRIMER, constrain $\fstar \in [0.014, 0.023]$ at $2\sigma$. At $\eta \geq 0.25$, absorbing $\sigm \geq 0.3$ requires $\fstar > 0.026$, exceeding this bound. At $\eta = 0.10$, even $\sigm = 5$ needs only $\fstar = 0.026$, marginally outside the $2\sigma$ range (Supplemental Material Table~S2, Fig.~S2). The UVLF combined with the SMF prior gives $\sigm \lesssim 0.3$\,cm$^2$\,g$^{-1}$ for $\eta \geq 0.25$, but no meaningful constraint for $\eta \lesssim 0.10$.

This degeneracy is not specific to SIDM. Any new physics producing a smooth, monotonic suppression of the UVLF will be similarly absorbed by the available astrophysical freedom, including warm dark matter, fuzzy dark matter, or modified gravity models. Constraints on such models from the UVLF shape alone should be treated with comparable caution unless the astrophysical parameters are independently anchored.

\begin{table}[h]
\centering
\caption{Astrophysical cost of absorbing the SIDM signal at
selected grid points. The profiled $\dchi$ is $< 0.02$ in all cases.
The SMF column indicates tension with the $2\sigma$ prior
$\fstar < 0.023$ from \cite{2024MNRAS.533.1808W}.}
\label{tab:cost}
\begin{tabular}{ccccc}
\toprule
$\sigm$ & $\eta$ & $\fstar^{\rm new}$ & $\Delta\sigUV$ & SMF \\
\midrule
1.0 & 0.10 & 0.023 & $+0.03$ & marginal \\
5.0 & 0.10 & 0.026 & $+0.03$ & $>2\sigma$ \\
0.5 & 0.25 & 0.026 & $+0.07$ & $>2\sigma$ \\
2.0 & 0.25 & 0.034 & $+0.07$ & $>2\sigma$ \\
\bottomrule
\end{tabular}
\end{table}

\section{Breaking the Degeneracy with Topology}
\label{sec:topology}

The reionization topology breaks the UVLF degeneracy because it probes a physically distinct channel from the same binding-energy modification. The duty-cycle enhancement depends on $\sigm$ through the blowout model, while the ionizing efficiency $\zeta$ is re-tuned at each $\sigm$ to maintain the target $\xHI$. Any adjustment to $\fstar$ or $\sigUV$ that hides SIDM in the UVLF therefore leaves the topology prediction unchanged. In semi-numerical simulations at $256^3$ resolution, the Euler characteristic of the ionization field increases by $(32 \pm 3)\%$ for $\sigm = 10$ relative to CDM, corresponding to $11\sigma$ significance from four independent density-field realizations \cite{Wang:2026uuo}.

\begin{figure}
\includegraphics[width=\columnwidth]{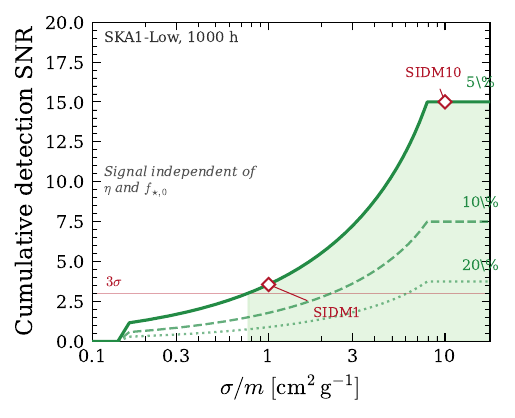}
\caption{Reionization topology as a nuisance-immune SIDM probe. Cumulative SKA1-Low detection SNR vs $\sigm$ at 5\% (solid), 10\% (dashed), and 20\% (dotted) systematic floors, in 1000\,h. The shaded region marks $> 3\sigma$ detection. Diamonds show the cross-sections simulated in \cite{Wang:2026uuo}. The topology signal is independent of $\eta$ and $\fstar$ at fixed $\xHI$.}
\label{fig:topology}
\end{figure}

Figure~\ref{fig:topology} shows the cumulative SKA1-Low detection significance as a function of $\sigm$. The $3\sigma$ detection threshold is $\sigm \approx 0.8$\,cm$^2$\,g$^{-1}$ at a 5\% systematic floor, $\approx 2.3$ at 10\%, and $\approx 6.1$ at 20\%. The forecast uses the analytic 21\,cm power spectrum ratio at fixed $\xHI$ \cite{Wang:2026uuo},
\begin{equation}
R(k) = \frac{b_{\gamma,\rm S}^2\,P_{mm}(k) + P_{\rm SN,S}}{b_{\gamma,\rm C}^2\,P_{mm}(k) + P_{\rm SN,C}},
\label{eq:P21ratio}
\end{equation}
where the subscripts S and C denote SIDM and CDM, $b_\gamma$ is the emissivity-weighted bias, and $P_{\rm SN} \propto 1/p$ is the shot-noise power. The dominant signal comes from the shot-noise term, which is suppressed by a factor of 1.8--3.3 as $p$ increases from 0.10 (CDM) to 0.18--0.30 (SIDM).

The two probes are complementary in the $(\sigm,\eta)$ parameter plane. Topology constrains $\sigm$ regardless of $\eta$ (a vertical band), while the UVLF+SMF constrains $\sigm$ as a function of $\eta$ (a diagonal). Because the probes use different instruments observing different sky areas at different wavelengths, their likelihoods factorize: $\mathcal{L}_{\rm joint} = \mathcal{L}_{\rm UVLF} \times \mathcal{L}_{\rm topo}$.

\begin{figure}
\includegraphics[width=\columnwidth]{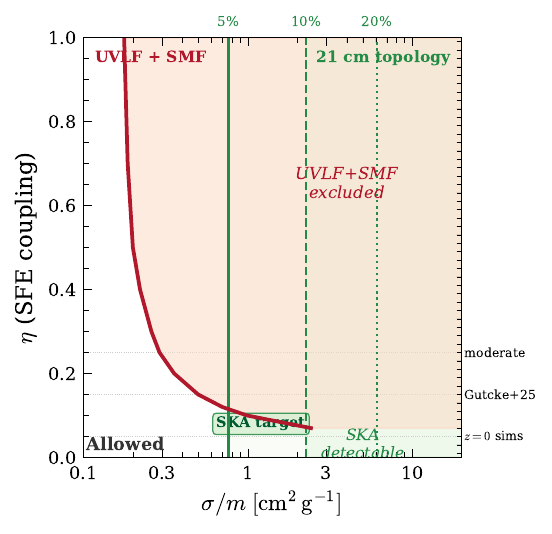}
\caption{Joint constraints in the $(\sigm,\eta)$ plane. The red curve shows the UVLF+SMF exclusion ($\fstar$ exceeds the $2\sigma$ SMF prior from \cite{2024MNRAS.533.1808W}). Green vertical lines mark SKA topology detection thresholds at 5\%, 10\%, 20\% systematic floors. The topology is independent of $\eta$, giving a vertical constraint. Right-axis ticks indicate $\eta$ values from SIDM simulations \cite{Robles_2017,Vogelsberger:2014pda,2025arXiv251005258G}.}
\label{fig:joint}
\end{figure}

Figure~\ref{fig:joint} shows the resulting joint constraint. Three regimes emerge. At $\eta \geq 0.25$, the UVLF+SMF excludes $\sigm \gtrsim 0.3$ while the topology detects $\sigm \gtrsim 0.8$ (5\% floor); both probes independently reject constant-cross-section SIDM above $\sigm \sim 0.8$, providing mutual confirmation even if one probe has uncontrolled systematics. At $\eta \sim 0.07$--$0.25$, the UVLF weakens but the topology retains sensitivity, defining a target region for SKA: SIDM in this window is hidden from the UVLF but visible in the 21\,cm morphology. An SKA non-detection here would push $\sigm$ below the values preferred by small-scale structure solutions, while a detection would constitute the first high-redshift evidence for SIDM. At $\eta \lesssim 0.07$, the UVLF provides no constraint and the topology alone determines the reach. This regime is favored by existing $z = 0$ SIDM simulations \cite{Robles_2017,Vogelsberger:2014pda}, which find negligible stellar mass changes in dwarf galaxies.

\begin{table}[h]
\centering
\caption{Joint constraints by $\eta$ regime. The UVLF+SMF column
gives the 95\% CL upper limit on $\sigm$ from the stellar mass
function prior. Topology thresholds are for SKA1-Low at 1000\,h.}
\label{tab:joint}
\begin{tabular}{lccc}
\toprule
$\eta$ range & UVLF+SMF & Topology & Status \\
\midrule
$\geq 0.25$ & $\sigm \lesssim 0.3$ & $\sigm \gtrsim 0.8$ &
  Both reject \\
$0.07$--$0.25$ & $\sigm \lesssim 0.3$--$2.4$ & $\sigm \gtrsim 0.8$ &
  SKA target \\
$\lesssim 0.07$ & None & $\sigm \gtrsim 0.8$ &
  Topology only \\
\bottomrule
\end{tabular}
\end{table}

\section{Discussion and Conclusions}
\label{sec:discussion}

The coupling parameter $\eta$ is the dominant theoretical uncertainty. At $z = 0$, SIDM simulations consistently find $\eta \lesssim 0.05$ at dwarf scales \cite{Robles_2017,Vogelsberger:2014pda,Correa:2020qam}, though Gutcke et al.\ \cite{2025arXiv251005258G} find $\eta \approx 0.10$--$0.15$ for velocity-dependent SIDM. At Milky Way masses, baryon contraction reverses the effect entirely \cite{2021MNRAS.507..720S}, indicating that $\eta$ is mass-dependent. No simulations exist at $z > 6$, where lower concentrations and shallower baryonic potentials could amplify the coupling. We adopt $\eta \in [0, 0.25]$ as a physically motivated range and note that dedicated high-redshift SIDM simulations with radiative transfer are needed to calibrate $\eta$ in the relevant regime (Supplemental Material \S E).

Velocity-dependent SIDM with large cross-sections at dwarf velocities and small ones at $v \gtrsim 100$\,km\,s$^{-1}$ can evade both probes, because the effective $\sigm$ drops between the topology-relevant and UVLF-relevant velocity scales (Supplemental Material \S D). Testing such models requires scale-dependent 21\,cm analyses exploiting the broad $k$-range of the power spectrum.

An additional complication is stochastic star formation. Spectroscopic observations at $z \sim 7$--$12$ reveal bursty star formation histories with SFR variations of $10$--$100\times$ within 100\,Myr \cite{2025ApJ...980..138H}. This burstiness inflates the effective $\sigUV$ and broadens the UVLF, mimicking one of the two channels through which the profile likelihood absorbs the SIDM signal. The BEACON survey further finds that UV-bright galaxies cluster more strongly than predicted by standard luminosity-halo mass relations\cite{kreilgaard2026beaconjwstnircampureparallel} , suggesting additional complexity in the SFE mapping that widens the astrophysical freedom beyond what our four-parameter model captures. Separating astrophysical burstiness from SIDM-induced effects may require the 21\,cm bispectrum or galaxy clustering statistics.

Several systematic uncertainties affect the quantitative thresholds but not the central result. The blowout exponent $\alpha_{\rm blow} = 0.7$ shifts the topology threshold by $\sim 30\%$ when varied over 0.5--1.0. The flat-core NFW approximation overestimates $\Dbind$ by $\sim 15\%$ relative to isothermal cores. The Sheth--Tormen HMF is uncertain at $z > 12$. The SMF prior from \cite{2024MNRAS.533.1808W} is measured at $z \sim 6$--$8$ and applied here to $z = 9$--$14$; if the SFE normalization evolves significantly across this range, the UVLF+SMF limits would shift accordingly. None of these systematics affect the degeneracy finding itself, which depends only on the smoothness of the SIDM UVLF suppression and the dimensionality of the astrophysical parameter space.

In summary: (1)~the UVLF alone cannot constrain SIDM, with profiled $\dchi = 0.002$--$0.21$ across 252 grid points; this degeneracy extends to any BSM physics producing smooth UVLF suppression; (2)~reionization topology provides a constraint on $\sigm$ independent of star formation astrophysics; (3)~the joint analysis excludes or detects constant-cross-section SIDM with $\sigm \gtrsim 1$--$2$ across the full $\eta$ range, establishing multi-probe approaches as essential for dark matter constraints at high redshift.

\section{Data Avaiability}
The code and data underlying this work are publicly available at \url{https://github.com/wzh800557-source/sidm-highz}. This includes the UVLF model framework, the 252-point profile likelihood scan results, verified JWST data points, figure generation scripts.

\begin{acknowledgments}

\end{acknowledgments}
\appendix
\section{SIDM binding energy model}
\label{sec:supp_binding}

\subsection{Thermalization radius}

The thermalization radius $r_1$ is defined by the condition that
the cumulative scattering optical depth reaches unity:
\begin{equation}
N_{\rm scat}(r_1) = \int_0^{t(z)} \rho_{\rm DM}(r_1)\,
\frac{\sigma_T}{m_\chi}\, v(r_1)\, dt = 1,
\end{equation}
where $\rho_{\rm DM}(r)$ follows an NFW profile with concentration
from Dutton \& Macci\`o (2014) and $t \simeq 0.35/H(z)$ is the
merger-corrected halo age (Fakhouri et al.\ 2010). The calibrated
scaling is
\begin{equation}
r_1 \simeq 0.8\;\kpc \left(\frac{\sigm}{1\;\mathrm{cm^2\,g^{-1}}}\right)^{0.55}
\left(\frac{M}{10^{10.5}\,\Msun}\right)^{0.35}
\left(\frac{8}{1+z}\right)^{0.5}.
\end{equation}

\subsection{Cored profile and binding energy}

The SIDM density profile is modeled as
$\rho_{\rm SIDM}(r) = \rho_{\rm NFW}(\max(r, r_1))$.
The gas binding energy within the star-forming region is
\begin{equation}
\Wg(< R) = \int_{r_{\rm inner}}^{R}
4\pi r^2 f_b \rho_{\rm DM}(r) |\Phi(r)| dr,
\end{equation}
with $f_b = 0.157$, $R = 0.1\,r_{\rm vir}$, and
$r_{\rm inner} = 0.3$\,kpc. The fractional reduction
$\Dbind = 1 - \Wg^{\rm SIDM}/\Wg^{\rm CDM}$ ranges from
$\sim 0.3$ ($\sigm = 1$) to $\sim 0.9$ ($\sigm = 10$) at
$M \sim 10^{10}$--$10^{11}\,\Msun$ and $z = 7$. Halos with
$r_1 < r_{\rm inner}$ are unaffected ($\Dbind = 0$).

\subsection{Two physical channels}

The binding-energy reduction drives two independent effects.
The duty cycle of ionizing-photon escape increases as
$p_{\rm SIDM} = p_{\rm CDM}(\Wg^{\rm CDM}/\Wg^{\rm SIDM})^{0.7}$,
enhancing the reionization topology. The star formation efficiency
decreases as $f_\star^{\rm SIDM} = f_\star^{\rm CDM}[1 - \eta\,\Dbind]$,
suppressing the UVLF.

\section{UVLF model and data}
\label{sec:supp_uvlf}

\subsection{Star formation efficiency}

The CDM baseline uses a double power-law SFE:
\begin{equation}
f_\star(M) = \frac{2\,\fstar}{(M/M_p)^{-\alow} + (M/M_p)^{0.5}},
\end{equation}
with $M_p = 10^{11}\,\Msun$ and redshift evolution
$\fstar(z) = \fstar(1+z)^{z_{\rm evol}}$. Each halo is assigned
$\mathrm{SFR} = f_\star f_b \dot{M}$ with $\dot{M}$ from
Fakhouri et al.\ (2010), converted to $M_{\rm UV}$ via the
Kennicutt (1998) relation, and scattered with Gaussian width
$\sigUV$. The UVLF is
$\Phi(M_{\rm UV},z) = \int (dn/d\ln M)\, p(M_{\rm UV}|M)\,d\ln M$
using the Sheth--Tormen HMF.

\subsection{CDM baseline}

\begin{figure}[h]
\includegraphics[width=0.95\linewidth]{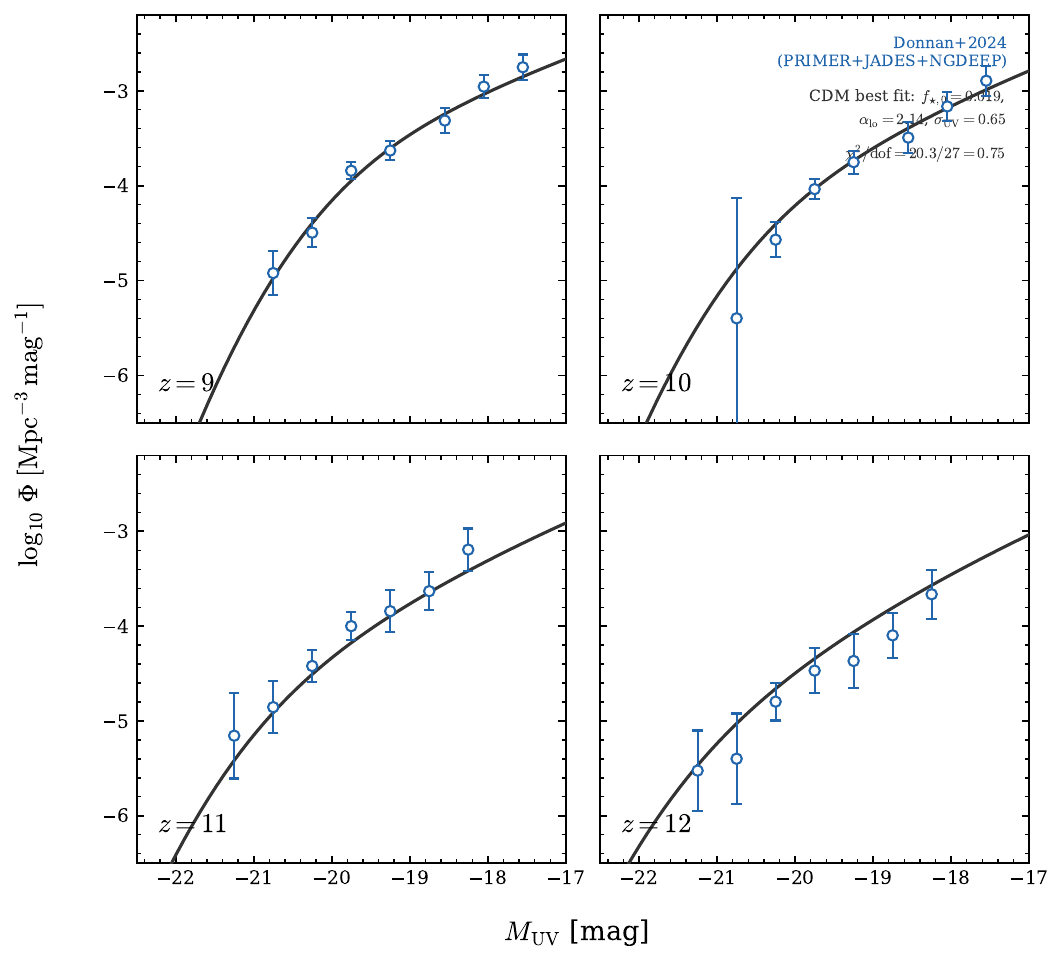}
\caption{CDM baseline fit to the JWST UVLF at $z = 9$, 10, 11, 12.
Data from Donnan et al.\ (2024). All 31 points fitted to within
$1.1\sigma$.}
\label{fig:supp_cdm}
\end{figure}

Minimizing $\chi^2$ over
$\bm{\theta}_{\rm astro} = \{\fstar,\alow,\sigUV,z_{\rm evol}\}$
at $\sigm = 0$ yields $\fstar = 0.019$, $\alow = 2.14$,
$\sigUV = 0.65$\,mag, $z_{\rm evol} = 0.13$, with
$\chi^2/\mathrm{dof} = 20.3/27 = 0.75$ (Figure~\ref{fig:supp_cdm}).

\section{Full profile likelihood results}
\label{sec:supp_profile}

\begin{table}[h]
\centering
\caption{Conditional $\dchi$ at fixed CDM astrophysics.}
\label{tab:supp_cond}
\begin{tabular}{ccccc}
\toprule
$\eta$ & $\sigm = 0.5$ & $\sigm = 1$ & $\sigm = 2$ & $\sigm = 5$ \\
\midrule
0.05 & 1.9 & 4.2 & 7.4 & 11.7 \\
0.10 & 8.1 & 17.9 & 32.1 & 51.7 \\
0.25 & 60.8 & 140.5 & 263.6 & 444.1 \\
0.50 & 334.2 & 836.1 & 1700.1 & 3093.1 \\
\bottomrule
\end{tabular}
\end{table}

\begin{table}[h]
\centering
\caption{Astrophysical cost of absorbing the SIDM signal. The last
column indicates tension with the SMF $2\sigma$ prior
$\fstar < 0.023$.}
\label{tab:supp_cost}
\begin{tabular}{cccccc}
\toprule
$\sigm$ & $\eta$ & $\Delta\fstar$ & $\fstar^{\rm new}$
& $\Delta\sigUV$ & SMF \\
\midrule
0.5 & 0.10 & $+12\%$ & 0.022 & $+0.03$ & OK \\
1.0 & 0.10 & $+19\%$ & 0.023 & $+0.03$ & marginal \\
2.0 & 0.10 & $+26\%$ & 0.024 & $+0.03$ & $>2\sigma$ \\
5.0 & 0.10 & $+34\%$ & 0.026 & $+0.03$ & $>2\sigma$ \\
\midrule
0.5 & 0.25 & $+33\%$ & 0.026 & $+0.07$ & $>2\sigma$ \\
1.0 & 0.25 & $+53\%$ & 0.030 & $+0.07$ & $>2\sigma$ \\
2.0 & 0.25 & $+77\%$ & 0.034 & $+0.07$ & $>2\sigma$ \\
5.0 & 0.25 & $+108\%$ & 0.040 & $+0.06$ & $>2\sigma$ \\
\bottomrule
\end{tabular}
\end{table}

\begin{figure}[h]
\includegraphics[width=\columnwidth]{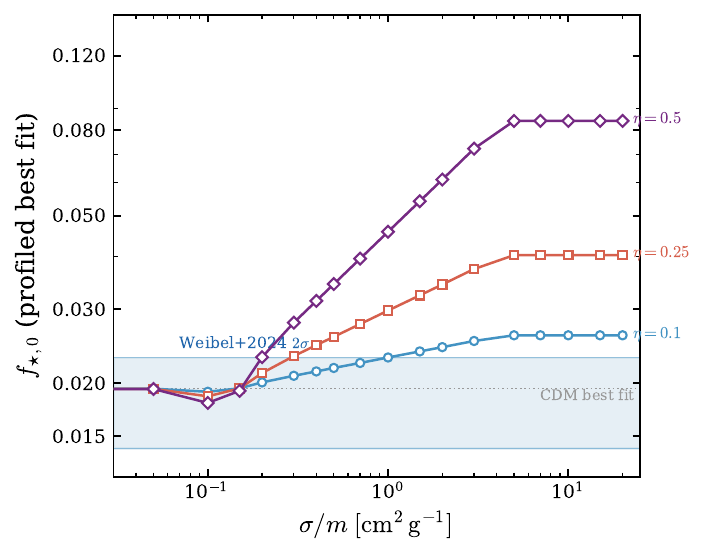}
\caption{Astrophysical cost: profiled $\fstar$ vs $\sigm$ at three
$\eta$ values. The blue band shows the $2\sigma$ SMF prior from
Weibel et al.\ (2024).}
\label{fig:supp_cost}
\end{figure}

\begin{figure}[h]
\includegraphics[width=\columnwidth]{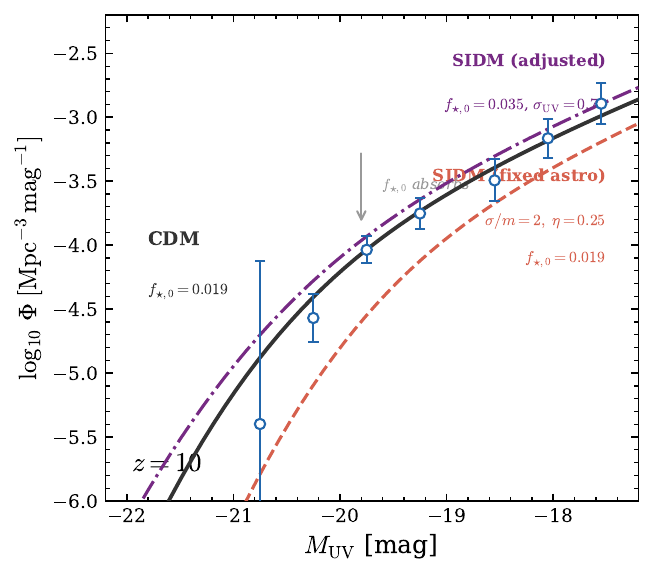}
\caption{UVLF at $z = 10$ showing CDM (solid), SIDM at fixed
astrophysics (dashed, $\sigm = 2$, $\eta = 0.25$,
$\dchi_{\rm cond} = 264$), and SIDM with re-optimized astrophysics
(dot-dashed, $\dchi_{\rm prof} = 0.018$). The arrow indicates the
$\fstar$ compensation.}
\label{fig:supp_uvlf_sidm}
\end{figure}

\section{Velocity-dependent SIDM}
\label{sec:supp_vdsidm}

\begin{figure}[h]
\includegraphics[width=\linewidth]{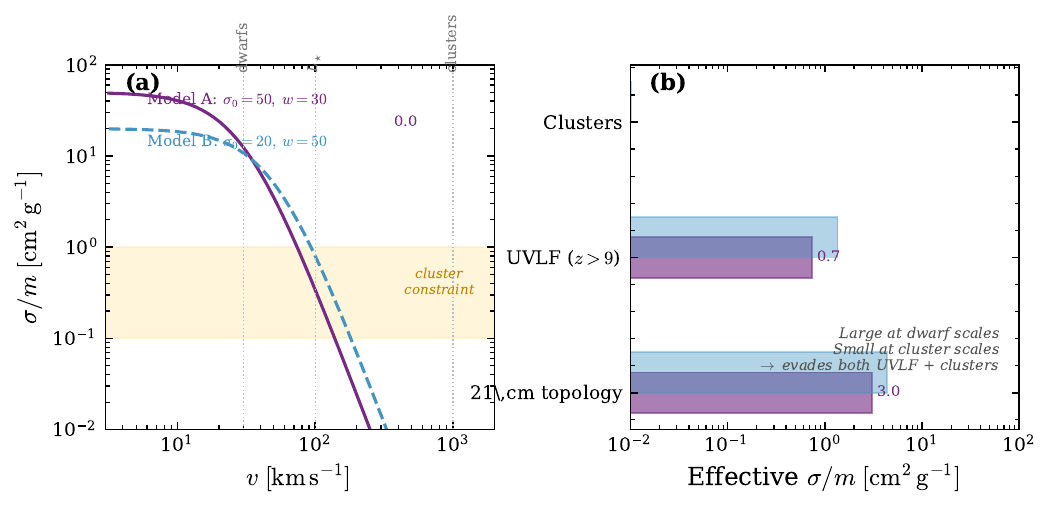}
\caption{Velocity-dependent SIDM. (a) Two Yukawa $\sigma(v)$ profiles.
(b) Effective $\sigm$ at each probe's velocity scale. Models with
large $\sigma_0$ and small $w$ maintain strong topology signals while
evading UVLF and cluster constraints.}
\label{fig:supp_vdsidm}
\end{figure}

Velocity-dependent models with
$\sigma(v) = \sigma_0/[1+(v/w)^2]^2$ can maintain large
cross-sections at dwarf velocities ($v \sim 30$\,km\,s$^{-1}$,
relevant for topology) while falling to
$\sim 1$\,cm$^2$\,g$^{-1}$ at
$v \sim 100$\,km\,s$^{-1}$ (relevant for the UVLF halo mass
range) and $\ll 1$ at cluster velocities. Such models evade both
the UVLF and topology constraints simultaneously
(Figure~\ref{fig:supp_vdsidm}), motivating scale-dependent 21\,cm
analyses.

\section{$\eta$ calibration from simulations}
\label{sec:supp_eta}

\begin{table}[h]
\centering
\caption{$\eta$ calibration from SIDM hydrodynamical simulations.}
\label{tab:supp_eta}
\begin{tabular}{lcccc}
\toprule
Reference & $M_{\rm halo}$ & $\sigm$ & $\Delta M_\star$ & $\eta$ \\
\midrule
Robles+17 & $10^{10}$ & 1 & $\sim 0\%$ & $\lesssim 0.05$ \\
Vogelsberger+14 & $10^{10}$ & 1,10 & $\sim 0\%$ & $\lesssim 0.05$ \\
TangoSIDM 24 & $10^{10\text{--}12}$ & v-dep & $\sim 0\%$ & $\sim 0$ \\
Gutcke+25 & LG dwarf & v-dep & $-25\%$ & $0.10$--$0.15$ \\
Sameie+21 & $10^{12}$ & 1,10 & $> 0$ & $< 0$ \\
\bottomrule
\end{tabular}
\end{table}

All existing simulations target $z = 0$. At $z > 6$, lower halo
concentrations and shallower baryonic potentials could amplify the
coupling. The baryon-contraction reversal seen at Milky Way masses
 is less likely at the lower masses and higher
redshifts relevant for our analysis. We adopt
$\eta \in [0, 0.25]$ as a physically motivated range.
\bibliography{apssamp}
\end{document}